\documentclass[11pt,twoside]{article} 
\usepackage{acta-info}
\usepackage{euler}
\usepackage{mathabx}

\newcommand{\vol}{\textbf{4,} 1 (2012) 7--16}   
\newcommand{\amss}{\amssubj{%
68N30
}}     
\newcommand{\CRs}{\CRsubj{%
F.3.1
}}   
\newcommand{\keyw}{\keywords{%
programming model, state space, program constructions, specification, correctness
}}   

\begin{document}

\title{%
Concept of the abstract program
}

\maketitle

\oneauthor{%
\href{http://people.inf.elte.hu/gt}{Tibor GREGORICS} 
}{%
\href{http://people.inf.elte.hu/gt}{E\"otv\"os Lor\'and University, Faculty of Informatics Budapest}
}{%
\href{mailto:gt@inf.elte.hu}{gt@inf.elte.hu}
}

\short{T. Gregorics}{Concept of the abstract program}

\begin{abstract}
The aim of this paper is to alter the abstract definition of the program of the theoretical programming model which has been developed at E\"otv\"os Lor\'and University for many years in order to investigate methods that support designing correct programs. The motivation of this modification was that the dynamic properties of programs appear in the model. This new definition of the program gives a hand to extend the model with the concept of subprograms while the earlier results of the original programming model are preserved.\end{abstract}


\noindent

\section {Introduction}

\medskip
It is a well-known aspiration to exclusively use methods which guarantee the correctness of the developed program with respect to the problem \cite{Dijkstra,Hoare,Gries} posed, that is, what makes it essential to find abstract mathematical definition of programming notions. At E\"otv\"os Lor\'and University, a relational model of programming which treats the most important, fundamental notions of programming---problem, program, state space, variable, data type, etc.---in a uniform and consistent way \cite{Fmodel,Fhun,Workgroup} has been built for thirty years.

The base of the abstract definitions of programming notions is the state space and their tool is the mathematical relation. The problem, for example, is a homogeneous binary relation over the state space that maps from the possible initial states to the appropriate goal states. The program is defined as all of its executions, so it can be described by the relation that maps from any state to the executions starting from one, where an execution is the sequence of the states concerned. The effect of a program is also a homogeneous binary relation where its domain contains the states from which the program surely terminates, i.e. the executions starting from these states are finite, and maps to states where these executions stop. The definition of the solution (when a program is said to solve a problem) establishes the main connection between the relations of the problem and the effect of the program. This theoretical approach makes it possible to investigate the concept and correctness of several kinds of program-designer methods such as program synthesis, analogous programming, etc. 

The main characteristic of this relational model is the special static point of view in which the concept of the program is a mapping and not a working, nonetheless it is originally dynamic. The advantage of this static point of view is the simple definition of the solution. Although it is not self-evident if the number of the components of the state space of a problem and that of a program are different. To prove the correctness, in this case, the program and the problem must be transformed into a new common state space \cite{Ftrans}. 

In the present model however, the state space is persistent, i.e. all variables of the program are global and static (the scope and life time of the variables involve the whole program), the so-called local variables cannot be created and destroyed during an execution. In absence of local variables, the concept of real subprograms cannot be introduced into our programming model. (Only macros can be defined.) Namely, the power of a subprogram is what permits creating new local auxiliary variables as well as parameter variables, and it is not supported by the present concept of the program. It is not a serious problem if simply the correctness is wanted to be proved but it is a great difficulty if our aim is to design a program. Unfortunately, the programming method, which is based on our programming model to synthetize correct programs, cannot produce subprograms with parameter variables, so their creation remains in the sphere of the implementation instead of the designing. It is an additional disadvantage that, without the concept of subprograms, recursive calling cannot be used in our programming model.

The aim of this paper is to remedy this shortcoming of the definition of the program, and to preserve the earlier results of the programming model.

\section {State space}

All definitions of the programming model are based on the state space. The concept of the state space has already been interpreted in several ways. For many people, the state space is a model of a von Neumann type  computer; others, e.g. Dijkstra \cite{Dijkstra}, associate this notion with the problem to be solved, where a state is a compound of the values of the main data types. So, the program is "outside" of the state space operating on it. In our programming model, this second meaning is used. In \cite{Fmodel}, the notion of the state space is a Cartesian product of the value set of data types. The only mistake of this obvious definition is that it imposes an order on the components. In \cite{Fhun}, where the \textit{\textbf{state space}} is a direct product, this mistake is repaired.

Here it is the definition of the direct product and the notions related to it.

Let $I$ be a finite set named as index set. Let $A_i$'s ($i \in I$) be arbitrary sets. The set
$\mathop{\bigtimes}\limits_{i \in I}{} A_i ::=\{x:I\rightarrow \bigcup_{i \in I} A_i \vert \forall i \in I : x(i) \in A_i  \}$ 
is the direct product of the sets $ A_i$  ($i \in I$).

The $A_i$'s are the components of the direct product $\mathop{\bigtimes}\limits_{i \in I}{} A_i$. The number of the components is finite. The direct product is the empty set if the set $I$ is empty. 

Let $A=\mathop{\bigtimes}\limits_{i \in I}{} A_i$ and $B=\mathop{\bigtimes}\limits_{j \in J}{} B_j$ be direct products where $I$ and $J$ are finite sets, $A_i$'s ($i \in
I$) and $B_j$'s ($j \in J$) are arbitrary sets. The $A$ and $B$ are equivalent if there exists a bijection $\nu:I \rightarrow J$ so that $\forall i \in I : A_i = B_{\nu(i)}$. In other words, B is the renamed A.

If $J \subseteq I$ and $\forall j \in J : B_j = A_j$ then $B$ is the subspace of $A$, i.e. $B \leq A$. 

The function $pr_B : A \rightarrow B$ is a projection if $B \leq A$ and $\forall a \in A : pr_B (a)= a \vert_J$. Not only one state but a set or sequence of states can also be projected if it is done one by one. If $J= \{ j \}$ then the function $pr_B$ is named as $j$ variable and $j$ is the variable name.

\section {The new concept of the program}

The most important notion of the programming model is the program. In our explanation, the program is not a collection of some statements that can be executed on a computer. The statements could only describe a program but the program is the complex of its executions. An execution is a sequence of states. A program, by definition, can always begin, i.e. at least one execution has to start from each state. If several executions start from the same start state, it means that the program is non-deterministic: nobody knows which execution will happen.

In the original programming model, every state of the progam belongs to the same state space. Now we are going to permit that the state space can be permanently changed; the inner states of the executions may have got new components.

The other novelty of our new definition is the idea of the  \textit{\textbf{base state space}} of the program. The first state (start state) of all executions and the last, if the execution is finite, are in this base state space.

Before giving the formal definition of the program, some notions must be introduced. Let $H^{**}$ denote the set of all finite and infinite sequences of the elements of set $H$. $H^{\infty}$  includes the infinite sequences; $H^*$ contains the finite ones. So, $H^{**} = H^* \cup H^{\infty}$ and $H^* \cap H^{\infty} = \emptyset$. The length of the sequence $\alpha \in H^{**}$ is $\left| \alpha \right|$, the value of which is $\infty$ if the sequence is infinite.

\begin{definition}
Let $A$ be the so-called base state space and $\bar{A}$  be the set of all states which belong to the state spaces $B$ whose subspace is $A$, i.e. $\bar{A}=\bigcup_{A \leq B} B$. The relation $S \subseteq A \bigtimes \bar{A}^{**}$ is a \textit{\textbf{program}} over $A$, if 
\begin{enumerate}
\item $\mathscr{D}_S = A$ \quad (domain of $S$),
\item $\forall \alpha \in \mathscr{R}_S \cap \bar{A}^* :  \alpha_{|\alpha|} \in A$ \quad  (the last state of the finite executions of
the range of $S$),
\item $\forall a \in A$ and $\forall \alpha \in S(a): |\alpha| \geq 1$ and  $\alpha_1=a$.
\end{enumerate}
\end{definition}

The variables of the base state space are the \textit{\textbf{base variables}}; the other variables are the \textit{\textbf{auxiliary variables}} of the program.

The concept of the program allows to create and destroy new components (variables) during an execution, so the state space changes dynamically. All new components have to be destroyed at the termination, at the very latest, but the base variables should never be removed. The current state always contains the components of the base state space.

The following three definitions will outline that the base state space of any program is denoted arbitrarily. The base variables of a program can be extended, restricted or renamed without changing the essence of the program.

The base variables of any program can be renamed easily without the execution of the program is changed. However, if the new name of a base variable would be identical to an auxiliary variable's name then this auxiliary variable is given a new, unique name.

\begin{definition}
Let the statespace $A$ and $B$ be equivalent, i.e. if $A=\mathop{\bigtimes}\limits_{i \in I}{} A_i$ and $B=\mathop{\bigtimes}\limits_{j \in J}{} B_j$ then there exists a bijection $\nu:I \rightarrow J$ so that $\forall i \in I : A_i = B_{ \nu (i)}$. Let $S$ be a program over the state space $A$. Let the name of the auxiliary variables of $S$ be $K$ ($K \cap I =  \emptyset$) and a bijection $\mu:K \rightarrow L$ so that $L \cap J =  \emptyset$. The program over $B$ is called the \textit{\textbf{variable renaming of S onto $\textit{B}$}} if its executions are identical to the executions of $S$, but in their all states, the variable name $i \in I$ is replaced by $\nu (i)$ and the variable name $k \in K$ is replaced by $\mu (k)$. 
\end{definition}

\vspace*{-6pt}
Not only can the name of the variables be changed but also the number of the components of the base state space: this state space can be extended with new variables, or some base variables can be degraded to auxiliary varibles.

The next definition shows that the base state space can be extended with a new component. The states of the executions can be extended with this new component if it is not the auxiliary variable of the original program. Otherwise if this component was an auxiliary variable, then it becomes a base variable; it should not be created and destroyed it, its life expands over the whole execution.

\begin{definition}
Let $S$ be a program over the state space $A=\mathop{\bigtimes}\limits_{i \in I}{} A_i$ and the variable name $k$  where $k \notin I$. (This $k$ may denote one of the auxiliary variables of $S$ or a totally new variable.) Let $C$ be the statespace $\mathop{\bigtimes}\limits_{i \in I \cup \{ k \} }{} A_i$. The \textit{\textbf{extension of S onto $\textit{C}$}} is the program (denoted by $S'$) whose base state space is $C$ and for all $c \in C:$
\end{definition}
\begin{tabbing}
 	$S'(c) = \{ \gamma\in \overline{C}^{**} \vert $ \= $\exists \alpha\in S(pr_A(c)) : |\alpha |=|\gamma| \wedge     
   \forall i \in [2..|\gamma|]:$ \\

	\>if  $k \notin \mathscr{D}_{\alpha_i }$  then $\gamma_i \vert_I = \alpha_i $  and $ \gamma_i (k) =  \gamma_{i-1}(k) $  \\ 
	\>if  $k \in \mathscr{D}_{\alpha_i }$ then $ \gamma_i = \alpha_i  \}$.
\end{tabbing}

It is easy to generalize this definition in case the state space $A$ is a subspace of the state space $C$.

The following definition shows how a base variable can be degraded to an auxiliary variable: the first step creates it and the last step destroys it.

\begin{definition}
Let $S$ be a program over the state space $A=\mathop{\bigtimes}\limits_{i \in I}{} A_i$ and let $C$ be the subspace of A. The \textit{\textbf{restriction of S onto $\textit{C}$}} is the program (denoted by $\overline{S}$) whose base state space is $C$ and for all $c \in C$: 
\end{definition}
\begin{tabbing}
	$\overline{S}(c)$ \= = \= \{ $<c, \alpha , pr_C(\alpha_{|\alpha|})> \in \overline{C}^* \vert \alpha \in S(a)\cap \overline{A}^*$ where $c=pr_C(a)  $ \} \\
	\> $ \cup $ \> \{ $< c, \alpha > \in \overline{C}^\infty  \vert  \alpha \in S(a)\cap \overline{A}^\infty$ where $c=pr_C(a)$ \}.
\end{tabbing}

The base state space of a program can be renamed, extended or restricted several times, thus the base state space can be totally transformed. But these transformations can change only the base state space and not the program. The essence of the executions of the program  remains the same. The renamed, extended and restricted program hardly differs from its original version. 

\begin{definition}
Two programs are \textit{\textbf{identical}} if they can transform into the same program, using extensions, restrictions and renaming.
\end{definition}

\section{Conclusions}

We have introduced a new concept of the program. Now let us observe what kinds of effects this modification has on the original programming model.

\subsection{Concept of dynamic program}

The new concept of the program smuggles the dynamic property of the program back to the static programming model \cite{Fmodel,Fhun} developed at E\"otv\"os Lor\'and University. Certainly, the fact that a program can create and destroy auxiliary variables is not a brilliant discovery. The difficulty of our investigation was to find out how this dynamic property can be embedded into the programming model described in the static point of view.

The most important element of our concept is the base state space. It can be seen as the interface of the program. It determines the components through which the program can communicate, and can make contact with its environment. Only the base variables can get value from outside before the program starts, and their value can be asked after the termination. 

The program becomes very flexible because its base state space can be changed easily. This property is reflected in the fact that the same program can possess different interfaces depending on the problem to be solved. The aim of a program can be changed by altering its base state space whereas its operation cannot be changed.

\subsection{Concept of solution}

Despite the above modification, the definition of the solution can be preserved \cite{Fmodel,Fhun,Workgroup}. The only thing that must be done is to redefine the concept of the \textit{\textbf{effect}} of the program because the definition of the solution relies on it.

Now we will repeat all definitions that are important to describe the concept of solution.

Let the problem generally be a relation $F\subseteq A\times A$ where $A$ is a state space. Let $S$ be a program over $A$. The relation $p(S)\subseteq A \bigtimes
 A$ is the effect of $S$ if 

\begin{enumerate}
\item $\mathscr{D}_{p(S)} = \{ a \in A \left| \right. S(a)\subseteq A^*  \} $
\item $\forall a \in \mathscr{D}_{p(S)}: p(S)(a) = \{ b \in A  \left| \right.  \exists \alpha \in S(a):  \alpha_{|\alpha|} = b \}$.
\end{enumerate} 

The program $S$ is said to solve the problem $F$ if 

\begin{enumerate}
\item $ \mathscr{D}_F \subseteq \mathscr{D}_{p(S)}$
\item $ \forall a \in \mathscr{D}_F: p(S)(a)\subseteq F(a)$
\end{enumerate}

The definition of the \textit{\textbf{solution}} supposes that the state space of the problem is identical to the base state space of the program. As the base state space can be chosen arbitrarily, the program can be extended, restricted or renamed in order that its base state space is identical to the state space of the problem. Since programs of this kind are identical, if one of them can solve the problem, so can the others.

Moreover, if the effects of the programs $S_1$ and $S_2$ are identical over all common base state spaces ($p(S_1) = p(S_2)$) and if one of them can solve a problem, so can the others. In this case these programs are called \textit{\textbf{equivalent}}. This relation is reflexive, symmetrical and transitive, thus it is an equivalence relation. Consequently, if one program belonging to an equivalence class can solve a problem, then every program derived from this class can also solve it.

In the original programming model, the definition of the solution must be generalized \cite{Ftrans} when the state spaces of the problem and the program are different. Now, our new concept is used to avoid this. It is enough to fit the base state space of the program to the state space of the problem and transform the program onto this common state space.

\vspace*{-3pt}
\subsection{Earlier results}

The original programming model contains many important results that can be used in the verification of programs. It is apparent that these results are not based directly on the concept of the program; except for the effect of the program. The definition of the effect of the program has not been changed because the new definition of the program fixes that the start state and the end state of the finite execution are in the same state space (that is, the base state space). Accordingly, all earlier results, namely Dijkstra's weakest precondition \cite{Dijkstra}, the theorem of the specification or the derivation rules \cite{Dijkstra,Fhun} of the program constructions \cite{Fhun} are used in unalterable forms.

Certainly, in the definitions of program constructions, the components (programs and conditions) which are the parts of the construction have to be on the same base state space. Accordingly, before making one of the constructions, this common base state space has to be agreed on.

\vspace*{-3pt}
\subsection{Subprograms}

The concept of real subprograms, which have got parameter variables, could not be introduced into the original programming model because the program has got an unvarying state space, thus local variables cannot be created during the execution, so parameter variables, which are local variables, cannot be used. Now, the situation is changed. In our new programming model, the concept of the subprograms may be defined. 

First and foremost the subprogram is a program. It can be executed independently; it has got own base state space. Its only speciality is that it can be built into another program. (This is the host program.) 

Each program, including subprograms, is equivalent to an assignment. This assignment is appropriate to identify the subprogram that is equivalent to it. Accordingly, in an arbitrary program description language, a subprogram can be denoted with this assignment. The variables of this assignment, which are the formal parameters, form the base state space of the subprogram. The variables at the left-hand side of the assignment are the output parameter variables; the ones at the right-hand side are the input parameter variables. (For simplicity's sake, we shall restrict our consideration to assignments where a parameter variable does not occur more than once.)

This assignment can be also used as a calling statement in the description of the host program. Certainly, in the calling statement, the parameter variables can be substituted by actual parameters (arguments), which are the expressions (often, they are only variables) of the host program. An output variable cannot be changed by an expression more general than a variable. The number and the type of the formal and current parameters must be the same. At this calling statement, the execution of the host program is interrupted, the control switches over the subprogram, the values of current parameters are given to the formal parameter variables, and the subprogram is executed. After the termination of the subprogram, the values of the output parameter variables are recopied to the appropriate current parameter variables. (Here, two kinds of parameter passing have been defined: passing by value and passing by value-result but other passing methods may be introduced.)

The base state space of the subprogram is the interface between the subprogram and the host program. At the beginning of the subprogram, the variables of this base state space will get their initial values from the current state of the host program. At the end of it these variables give their values to the variables of the host program. The parameter variables and other local variables of the subprogram are created when the subprogram is called and destroyed at the termination of the subprogram.

In addition, the usage of subprograms permits recursive callings because a subprogram can call itself. After each calling, the parameter and the local variables of the subprogram are created again and again without the error message ''out of memory'' because the memory of the abstract program is infinite.

During planning, it is convenient that all the current variables of the host program are treated as global with respect to the subprogram that is called by the host program. The applied programming style and the facilities of the selected programming language determine if these global variables can be used directly in the subprogram or not. Obviously, their use has to be forbidden if we want to guarantee the independency of the subprogram. 

We can make difference between the calling statement and the calling expression. In the latter case, only the right-hand side of the subprogram's head appears inside an expression of the host program, which contains actual parameters instead of formal parameter variables. After the termination, the result value of the left-hand side of the head is given back to the place of the calling expression in the host program.

Certainly, the abstract program description language which is used for planning has to be extended with the notation of the subprogram and of its calling including the connection between the actual parameters and the parameter variables.

To sum up the introduction of the concept of the subprograms makes it possible to design subprograms during planning and not only during implementation.  

\vspace*{-3pt}
\section*{Acknowledgements}
This paper is supported by the European Union and co-financed by the European Social Fund (grant agreement no. T\'AMOP 4.2.1/B-09/1/KMR-2010-0003).

I would like to thank \'Akos F\'othi  for his pieces of useful advices.

\bigskip
\rightline{\emph{Received: March 14, 2012 {\tiny \raisebox{2pt}{$\bullet$\!}} Revised: April 14, 2012}} 
 
\end{document}